\documentclass[11pt]{article}
\date{\today}
\usepackage[usenames,dvipsnames]{color}
\usepackage{amsmath,amsthm,amsfonts,amssymb,multicol,amscd,amsbsy}
\usepackage{graphicx}
\usepackage{booktabs}
\usepackage{array}
\usepackage{subfigure}
\usepackage{enumerate}
\usepackage{url}
\usepackage[nodayofweek]{datetime}
\usepackage[english]{babel}
\usepackage{mathtools}
\usepackage[toc,page]{appendix}
\usepackage{verbatim} 
\usepackage{amsthm} 
\usepackage{enumitem}
\usepackage{enumerate}
\usepackage{bbm} 
\usepackage[normalem]{ulem} 
\usepackage{bm}
\usepackage{hyperref}
\usepackage{latexsym}
\usepackage{color}

\DeclarePairedDelimiter\ceil{\lceil}{\rceil}
\DeclarePairedDelimiter\floor{\lfloor}{\rfloor}
\newcommand{\fl}[1]{\left \lfloor{#1}\right \rfloor}

\usepackage{authblk}

\newcommand{\be}{\begin}
\newcommand{\e}{\end}
\newcommand{\beq}{\begin{equation}}
\newcommand{\eeq}{\end{equation}}
\newcommand{\beqs}{\begin{equation*}}
\newcommand{\eeqs}{\end{equation*}}

\newcommand{\ul}{\underline}

\renewcommand{\l}{\left}
\renewcommand{\r}{\right}
\renewcommand{\d}{\mathrm{d}} 
\newcommand{\De}{\Delta}
\renewcommand{\Re}{\mathrm{Re}}

\newcommand{\set}[1]{\mathbb{#1}}
\newcommand{\curly}[1]{\mathcal{#1}}

\newcommand{\setof}[2]{\left\{ #1\; : \;#2 \right\}}

\newcommand{\R}{\set{R}}
\newcommand{\C}{\set{C}}
\newcommand{\Z}{\set{Z}}
\newcommand{\E}{\set{E}}
\newcommand{\T}{\set{T}}

\newcommand{\ev}{\mathbf{e}}
\newcommand{\od}{\mathbf{o}}

\newcommand{\om}{\omega}
\newcommand{\Om}{\Omega}
\newcommand{\eps}{\epsilon}

\newcommand{\lam}{\lambda}

\newcommand{\al}{\alpha}
\newcommand{\de}{\delta}

\newcommand{\dist}{\mathrm{dist}}

\newcommand{\ind}{\mathbbm{1}}		

\newcommand{\ttmatrix}[4]{\left(\be{array}{cc} #1&#2\\	#3&#4 \e{array}	\right)}

\newcommand{\Tr}{\mathrm{Tr}}	

\newtheorem{thm}{Theorem}[section]
\newtheorem{lm}[thm]{Lemma}
\newtheorem{cor}[thm]{Corollary}

\theoremstyle{definition}

\numberwithin{equation}{section}

\theoremstyle{remark}

\def\dotuline{\bgroup
  \ifdim\ULdepth=\maxdimen  
   \settodepth\ULdepth{(j}\advance\ULdepth.4pt\fi
  \markoverwith{\begingroup
  \advance\ULdepth0.08ex
  \lower\ULdepth\hbox{\kern.15em .\kern.1em}%
  \endgroup}\ULon}

\def\dashuline{\bgroup
  \ifdim\ULdepth=\maxdimen  
   \settodepth\ULdepth{(j}\advance\ULdepth.4pt\fi
  \markoverwith{\kern.15em
  \vtop{\kern\ULdepth \hrule width .3em}%
  \kern.15em}\ULon}
\allowdisplaybreaks

\begin{document}

\title{Weyl sums and the Lyapunov exponent for the skew-shift Schr\"odinger cocycle}
\author[1,2]{Rui Han}
\author[1,3]{Marius Lemm}
\author[1,4]{Wilhelm Schlag}
\affil[1]{\small{School of Mathematics, Institute for Advanced Study}}
\affil[2]{\small{Department of Mathematics,  Georgia Institute of Technology}}
\affil[3]{\small{Department of Mathematics,  Harvard University}}
\affil[4]{\small{Department of Mathematics,  Yale University}}
\maketitle

\abstract{We study the one-dimensional discrete Schr\"odinger operator with the skew-shift potential $2\lam\cos\l(2\pi \l(\binom{j}{2} \omega+jy+x\r)\r)$. This potential is long conjectured to behave like a random one, i.e., it is expected to produce Anderson localization for arbitrarily small coupling constants $\lam>0$. In this paper, we introduce a novel perturbative approach for studying the zero-energy Lyapunov exponent $L(\lam)$ at small $\lam$. Our main results establish that, to second order in perturbation theory, a natural upper bound on $L(\lam)$ is fully consistent with $L(\lam)$ being positive and satisfying the usual Figotin-Pastur type asymptotics $L(\lam)\sim C\lam^2$ as $\lam\to 0$. The analogous quantity behaves completely differently in the Almost-Mathieu model, whose zero-energy Lyapunov exponent vanishes for $\lam<1$. The main technical work consists in establishing good lower bounds on the exponential sums (quadratic Weyl sums) that appear in our perturbation series. }

\section{Introduction and main results}
A central task of ergodic theory is to compare the orbits of a given ergodic dynamical system with sequences of i.i.d.\ random variables. For instance, we can phrase the classical ergodic theorem as the statement that empirical means along orbits are asymptotically indistinguishable from empirical means of independent and identically distributed (i.i.d.) random variables distributed according to the equilibrium measure. Going beyond the ergodic theorem, refined comparisons to the random case typically involve \emph{correlations} within the sequences. For example, it is known that certain ergodic dynamical systems exhibit the Poissonian two-point correlations and Poissonian spacing associated with i.i.d.\ sequences; see, e.g., \cite{RS,RSZ} for ergodic systems related to the skew-shift. 

The comparison between the orbits of a dynamical system and an i.i.d.\ sequence can also be made from the perspective of a \emph{quantum particle living on $\Z$}. In a nutshell, the question becomes whether the orbits are sufficiently ``random-like'' to localize the quantum particle in a finite region of space. Localization occurs due to destructive interference of waves and therefore it depends crucially on correlations within the underlying dynamical system. 

Let us now define the model precisely. We introduce the Schr\"odinger operator $-\De+\lam v$ on $\ell^2(\Z)$ whose parameters space consists of the real-valued sequence of ``potentials'' $v=\{v_j\}_{j\in\Z}$ and the global ``coupling constant'' $\lam>0$.  By definition, the Schr\"odinger operator maps a sequence $\psi\in \ell^2(\Z)$ to the sequence
$$
((-\De+\lam v)\psi)_j:=\psi_{j+1}+\psi_{j-1}+\lam v_j \psi_j.
$$
The basic idea is then to generate the bi-infinite sequence of potentials $v=\{v_j\}_{j\in\Z}$ by sampling along the orbits of an underlying ergodic dynamical system, and to compare the resulting Schr\"odinger operator $-\De+\lam v$ with one that is generated by an i.i.d. sequence of $\{v_j\}_{j\in\Z}$.

We first recall the benchmark for ``random-like behavior'' of these models, i.e., we take $\{v_j\}_{j\in\Z}$ to be an i.i.d.\ family of random variables. In that case, the Schr\"odinger operator exhibits \emph{Anderson localization} \cite{Anderson} for any $\lam>0$. This means, for instance, that its eigenfunctions decay exponentially \cite{KS}; for further references, see \cite{AW}.

Anderson localization, specifically, the exponential decay of Schr\"o\-dinger eigenfunctions is closely related to the associated cocycle having a positive \emph{Lyapunov exponent} for any $\lam>0$ \cite{Fur}. Let us recall the definition of the Schr\"odinger cocycle and the associated Lyapunov exponent $L(\lam,E)$. Consider a general Schr\"odinger operator $-\De+\lam v$ on $\ell^2(\Z)$ whose (real-valued) sequence of potentials $\{v_j\}_{j\in\Z}$ is generated by some underlying dynamical system. The eigenvalue equation reads $(-\De+\lam v)\psi=E\psi$ on $\ell^2(\Z)$, with $E\in\R$. It is a second-order difference equation and can therefore be solved by iteratively applying transfer matrices $A_j$, given by
\beq\label{eq:Ajdefn}
\begin{aligned}
A_j:=\ttmatrix{E-\lam v_j}{-1}{1}{0}.
\end{aligned}
\eeq
Since the transfer matrices $A_j$ depend on the orbit of the underlying dynamical system through $v_j$, they generate a cocycle and we can define the associated Lyapunov exponent $L(\lam,E)$ via
\beq\label{eq:Ldefn}
L(\lam,E):=\lim_{n\to\infty} \frac{1}{n}\log \Tr[M_n^* M_n],\qquad \textnormal{with }\, M_n:=A_n \ldots A_1.
\eeq
The limit exists by the F\"urstenberg-Kesten theorem or Kingman's subadditive ergodic theorem, under appropriate assumptions on the underlying dynamical system \cite{Viana}.

If the Lyapunov exponent $L(\lam,E)$ is \emph{strictly positive} at an energy $E$ which lies in the spectrum of the Schr\"odinger operator, then this strongly indicates (but does not imply) that the model exhibits localization and therefore ``random-like behavior'' at that energy $E$. In this paper, we therefore focus on the positivity of the Lyapunov exponent as the telltale sign of localization.\\

Before we introduce the skew-shift potential, let us consider the most natural ergodic system --- circle rotation (or a shift on the $1$-torus). First, for periodic sequences of $\{v_j\}_{j\in\Z^d}$, the Lyapunov exponent vanishes everywhere inside the spectrum. (E.g., when $v_j=0$, the eigenfunctions are plane waves.) In other words, the lack of ergodicity of \emph{rational} circle rotation fails to localize the quantum particle. 

For rotation by an irrational angle, the situation changes. This is the case of the famous Almost-Mathieu operator, whose potential is given by
\beq\label{eq:AMO}
v_j=2\cos(2\pi(j\al+\theta)),
\eeq
with $\al\in [0,1]\setminus\mathbb Q$ and $\theta\in [0,1]$.  In this case, the positivity of the Lyapunov exponent depends critically on the size of the coupling constant $\lam>0$. We have the bound $L(\lam,E)\geq \log \lam$ by Herman's subharmonicity trick \cite{Her}, so $L(\lam,E)>0$ for $\lam>1$ at all energies. The threshold $\lam=1$ is sharp, i.e., 
$L(\lam,E)=0$ for $0<\lam<1$, and $E$ in the spectrum of the Schr\"odinger operator. (This follows from the duality properties of the model under Fourier transformation.) For later, we note that $E=0$ is in the spectrum \cite{BelSim,BJ} and so $L(\lam,0)=0$ when  $0<\lam<1$ and $v_j$ is given by \eqref{eq:AMO}. To summarize, the weak ergodic properties of irrational circle rotation are only sufficient to localize the quantum particle subjected to \eqref{eq:AMO}, \emph{if} the coupling constant $\lam$ is sufficiently large.

\medskip

In this paper, we consider an ergodic potential which is believed to be ``slightly more random'' than \eqref{eq:AMO}. It is obtained by projecting orbits of the standard skew-shift on the $2$-torus on its first coordinate, leading to the potential
\beq\label{eq:skewshiftpotential}
v_j(x,y)=2\cos\l(2\pi \l(\binom{j}{2} \omega+jy+x\r)\r),
\eeq
where $\omega\in[0,1]\setminus \mathbb Q$ and $x,y \in [0,1]$ are parameters. (We call $\om$ the ``frequency''.) 

The key difference between \eqref{eq:skewshiftpotential} and \eqref{eq:AMO} is the appearance of a quadratic term, $j^2\om$. Rudnick, Sarnak, and Zaharescu \cite{RSZ} conjectured that the fractional part of such sequences exhibits Poissonian spacing (and proved that this holds for topologically generic $\om$ along a subsequence of $j\to\infty$). The phenomenon of Poissonian spacing also occurs for i.i.d.\ sequences, but not for the fractional parts of $j\om$ (i.e., not for circle rotation) which, by contrast,  exhibits level repulsion \cite{Ble,PBG}. Other results in this direction were proved in \cite{DRH,MS,MY,RS}.

The conjecture that \eqref{eq:skewshiftpotential} is ``more random-like'' than \eqref{eq:AMO} from the perspective of a quantum particle can now be phrased as follows.

\be{conj}\label{conj}
For the potential \eqref{eq:skewshiftpotential}, one has $L(\lam,E)>0$ for all $\lam>0$ and all $E\in\R$.
\e{conj}

We note that Herman's subharmonicity trick, which holds in a wider context \cite{SoSp}, also applies to the Schr\"odinger operator with $v_j$ given by \eqref{eq:skewshiftpotential}. It still implies that $L(\lam,E)\geq \log \lam$ is positive for $\lam>1$, so Conjecture \ref{conj} is only concerned with $0<\lam\leq 1$. 

The Schr\"odinger operator with skew-shift potential has been studied in \cite{BGS} using the large deviation approach to Lyapunov exponents \cite{B1,BG,GS}. In \cite{BGS}, Anderson localization was derived for large $\lam$; see also the recent effective version \cite{HLS}. So far, however, there has been little concrete evidence for Conjecture \ref{conj}, i.e., for random-like behavior of the skew-shift potential with $0<\lam<1$. We are only aware of a work by Bourgain \cite{B2} which studies the closely related potential $v_j=2\cos(2\pi j^2\om)$, and an unpublished preprint by  Kr{\"u}ger \cite{Kru}. The former establishes that, for any $\lam>0$ and a positive-measure set of frequencies $\om$, the Schr\"odinger operator has point spectrum whose closure has positive measure. The latter establishes the positivity of the Lyapunov exponent for a modified skew-shift model.

\medskip

In this paper, we make a modest first contribution towards Conjecture \ref{conj}. We approach the problem perturbatively, i.e., we consider the zero-energy Lyapunov exponent $L(\lam,0)$ as a power series in $\lam>0$.  (In analogy with the random case, it is expected that for the skew-shift the spectrum is an interval. In particular, $E=0$ should be in the interior of the spectrum.) We study a natural upper bound on $L(\lam,0)$ which is obtained by Jensen's inequality and which we expect captures some of the essential features, as we formulate in Conjecture \ref{conj'}. We hope that our results motivate further research into the delicate localization question for the skew-shift model.

To a large extent, the motivation for this work stems from our earlier paper~\cite{HLS} where the positivity of the Lyapunov exponent was derived from finite-volume properties at a sufficiently large scale. One of these properties is the growth of $\| M_n\|$ as $n\to\infty$, at least generically in the phase parameters. We therefore average the trace in \eqref{eq:Ldefn} without first taking a logarithm. 
We show that this alternative conjecture is true to second order in perturbation theory. In fact, the result is consistent with the Figotin-Pastur asymptotics $L(\lam,0)\sim c\lam^2$ as $\lam\to 0$. (The Figotin-Pastur asymptotics are expected to hold if Conjecture \ref{conj} is true.)

\medskip
  
We now summarize the main contributions of this paper.

\be{itemize}
\item Our first main result, Theorem \ref{thm:main}, provides manageable formulae for the two lowest-order coefficients of the relevant perturbation series (cf.\ the original formula in Proposition \ref{prop:pertseries}). The first-order term in the perturbation series can be computed directly and we see that it behaves markedly differently from the Almost-Mathieu case. The second-order term is given by a sum over quadratic Weyl sums. The growth properties of these exponential sums are well--known to be related to  questions in number theory, specifically about the number of solutions to Diophantine equations.  
\item 
In our second and third main results, Theorems \ref{thm:main2} and \ref{thm:main3}, we prove complementary lower bounds on the relevant Weyl sums. (They are complementary in the order of quantifiers.) These are the key results on a technical level. Both results are ultimately based on rational approximation and asymptotic formulae in the spirit of Hardy and Littlewood, but the details are quite different. Theorem \ref{thm:main2} is proved by a probabilistic argument (second moment method), with input from the central limit theorem for purely quadratic Weyl sums proved in \cite{JvH}. Theorem \ref{thm:main3} is based on asymptotic formulae for frequencies that are close to a rational and a variant of Khinchin's theorem \cite{FJK}.

\item Taking a clue from the cluster expansion method from statistical mechanics, we rephrase Conjecture \ref{conj} as a counting problem, namely, as a precise relation between the number of solutions to certain Diophantine equations (Section \ref{sect:diophantine}).
\e{itemize}

The paper is organized as follows. In Section 2, we present our main results Theorems \ref{thm:main}, \ref{thm:main2} and \ref{thm:main3}. These are proved in Sections 3--5, respectively. In Appendix A, we derive a similar perturbation series for the Almost-Mathieu model, and in Appendix B, we discuss the relation between our setup and the one studied recently using homogeneous dynamics \cite{Cel,CM,Marklof}.

\paragraph{Acknowledgments}
The authors are grateful to the Institute for Advanced Study for its hospitality during the 2017-2018 academic year. They thank I.~Jauslin, P.~Sarnak and T.~Spencer for useful discussions. This material is based upon work supported by the National Science Foundation under Grant No.~DMS-1638352. The third author was partially supported by the NSF through DMS-1500696.

\section{Main results}
\subsection{Setup}
Given a function $f:[0,1]^2\to \R$, we introduce the notation
$$
\mathbb E_{\mathbb T^2}[f]:=\int_0^1 \int_0^1 f(x,y) \d x \d y.
$$
For the skew-shift potential \eqref{eq:skewshiftpotential} at irrational frequency $\om$, the F\"ursten\-berg-Kesten theorem implies that the Lyapunov exponent \eqref{eq:Ldefn} can be computed by the following spatial average
$$
L(\lam,E)=\lim_{n\to\infty} \frac{1}{n} \mathbb E_{\mathbb T^2}\l[\log \Tr[M_n^*M_n]\r].
$$
Hence, by Jensen's inequality, we have the upper bound
$$
L(\lam,E)\leq \liminf_{n\to\infty} \frac{1}{n} \log \mathbb E_{\mathbb T^2}\l[\Tr[M_n^*M_n]\r].
$$
We see that a \emph{necessary condition} for Conjecture \ref{conj} to hold is that the following polynomial in $\lam$,
\beq\label{eq:Pndefn}
\curly{P}_n(\lam,E):=\mathbb E_{\mathbb T^2} \l[\Tr[M_n^* M_n]\r],
\eeq
satisfies the following variant of Conjecture \ref{conj}.
 
\be{conj}\label{conj'}
There exists $c_\lam>0$ such that
\beq\label{eq:necessary}
\curly{P}_n(\lam,E)\geq \exp(n (c_\lam+o(1))),
\eeq
as $n\to\infty$. 
\e{conj}

 In fact, taking a clue from the Figotin-Pastur asymptotics established for the Lyapunov exponent of potentials with better ergodic properties \cite{FigPas,CS}, we expect that 
 \beq\label{eq:FP}
 c_\lam=C\lam^2+o(\lam^2),
 \eeq
 for some constant $C>0$, as $\lam\to 0$.\\

We focus on the zero-energy case, $\curly{P}_n(\lam,0)$, and establish Conjecture \ref{conj'} to second order in perturbation theory in $\lam$. More specifically, the polynomial $\curly{P}_n(\lam,0)$ is an even function of $\lam$ and we show that its $\lam^2$ and $\lam^4$ coefficients are consistent with \eqref{eq:necessary} and \eqref{eq:FP}. That is, we show that $\curly{P}_n(\lam,0)\geq 2n\lam^2+\frac{C}{2}n^2\lam^4+\ldots$ for some constant $C>0$.

 We remind the reader that at $E=0$ and $0<\lam<1$ the Lyapunov exponent of the Almost-Mathieu model \eqref{eq:AMO} \emph{vanishes}. In fact, one can easily see that also the polynomial $\curly{P}_n(\lam,0)$ behaves completely differently in the Almost-Mathieu case --- the coefficient of $\lam^2$ remains bounded in $n$. (See Appendix \ref{app:AMO}.)
 
 This shows that the application of Jensen's inequality above apparently did not destroy the critical difference between these two models for small $\lam$. Consequently, our results indicate that the Lyapunov exponent for  \eqref{eq:AMO} and \eqref{eq:skewshiftpotential} behave markedly differently for small $\lam$. 

\subsection{The perturbation series}
We have found it suitable to take a direct approach to the perturbation theory. This is in contrast to the perturbation theory for the Lyapunov exponent successfully used by Figotin-Pastur \cite{FigPas} in the random case and Chulaevsky-Spencer \cite{CS} for some deterministic potentials. 

It turns out that the zero-energy condition leads to certain \emph{parity conditions} on the summands. This phenomenon significantly complicates matters and does not usually appear either in the context of perturbation theory in statistical mechanics, or in the context of exponential sums.

We denote the set of even (odd) integers by $\mathbf{e}$ and $\mathbf{o}$, respectively. Let $\ul{j}=(j_1,\ldots,j_{k_1})$ and $\ul{l}=(l_1,\ldots,l_{k_2})$ be two vectors with integer entries. The set $\clubsuit$ is defined by the conditions
\beq\label{eq:clubdefn}
\begin{aligned}
(\ul{j},\ul{l})\in \clubsuit
\,\Longleftrightarrow\, 
j_1-l_1\in \mathbf{e},\quad&\textnormal{ and }\quad j_{s+1}-j_s\in\mathbf{o},\quad\forall 1\leq s\leq k_1-1,\\
&\textnormal{ and }\quad
\,l_{s+1}-l_s\in\mathbf{o},\quad \forall 1\leq s\leq k_2-1.
\end{aligned}
\eeq
Recall Definition \eqref{eq:Pndefn} of $\curly{P}_n(\lam,E)$.

\be{prop}\label{prop:pertseries}
Let $\om\in[0,1]$ and let $n\geq 1$ be an integer. The polynomial
$$
\curly{P}_n(\lam,0)
=\sum_{k=0}^{n} \al_{2k} \lam^{2k}
$$
has coefficients given by $\al_0=2$ and 
\beq\label{eq:al2kdefn}
\al_{2k}=\sum_{\substack{0\leq k_1,k_2\leq  n\\ k_1+k_2=2k\\ k_1-k_2\equiv 0 \mod 4}} 
\sum_{\substack{1\leq j_1<\ldots<j_{k_1}\leq n\\ 1\leq l_1<\ldots<l_{k_2}\leq n}} \ind_{\clubsuit}(\ul{j},\ul{l}) \mathbb E_{\mathbb T^2}[ v_{j_1}\ldots v_{j_{k_1}} v_{l_1}\ldots v_{l_{k_2}}],
\eeq
for $k\geq 1$.
\e{prop}

Let us also denote
$$
e[x]:=\exp(2\pi i x),\qquad c[x]:=\cos(2\pi x),
$$
and write $\ul{1}$ for the vector $(1,\ldots,1)$ of a length that is given from context. The expectation appearing in \eqref{eq:al2kdefn} can be expressed as an exponential series
\beq\label{eq:expectation}
\begin{aligned}
\mathbb E_{\mathbb T^2} [v_{j_1}\ldots v_{j_{k_0}}]
=&\Re \sum_{\substack{\ul{a}\in\{\pm 1\}^{k_0}\\ \ul{a}\perp \ul{1}, \ul{j}}}  e\l[\frac{\om}{2}\sum_{s=1}^{{k_0}} a_s j_s^2\r]\\
=&2\sum_{\substack{\ul{a}\in\{\pm 1\}^{k_0}\\ a_1=1\\ \ul{a}\perp \ul{1}, \ul{j}}} c\l[\frac{\om}{2} \sum_{s=1}^{{k_0}} a_s j_s^2\r],
\end{aligned}
\eeq
(For the second equality, we used that $c$ is an even function to fix $a_1=1$.) 

Above, we used the notation $\ul{a}\perp \ul{1}, \ul{j}$ to encode the equations
$$
\sum_{s=1}^{k_0} a_s=0,\qquad \sum_{s=1}^{k_0} a_s j_s=0.
$$

\be{rmk}
We point out that the orthogonality condition $\ul{j}\perp \ul{1}$ in \eqref{eq:expectation} is a consequence of the fact that the potential \eqref{eq:skewshiftpotential} is generated by the true skew-shift. It would be absent for the potential $v_j=2\cos(2\pi j^2\om)$, for instance. 
\e{rmk}

\subsection{Identities for $\al_2$ and $\al_4$}
Our first main result, Theorem \ref{thm:main}, concerns the lowest order coefficients, $\al_2$ and $\al_4$, which are a priori defined by the rather unwieldy formula \eqref{eq:al2kdefn}. For small $\lam$, 
Conjecture~\ref{conj'} translates to the lower bounds
\beq\label{eq:necessary'}
\al_2\geq Cn,\qquad \al_4\geq C\frac{n^2}{2},
\eeq
which should hold for some constant $C>0$, as $n\to\infty$.

\be{thm}[First main result]\label{thm:main}
Let $\om\in [0,1]$. For any integer $n\geq 1$, we have
\begin{align}
\label{eq:al2}
\al_2=&2n,\\
\label{eq:al4}
\al_4=& 4\sum_{m=1}^{\fl{\frac{n}{2}}} 
\l|\sum_{l=1}^m  \:  e\l[\om\l(l^2-l\r)\r]\r|^2
+4\sum_{m=1}^{\fl{\frac{n-1}{2}}}\l|\sum_{l=1}^m  \:  e\l[\om\l(l^2-l\r)\r]\r|^2.
\end{align}
\e{thm}

\be{rmk}
\be{enumerate}[label=(\roman*)]
\item
Notice that Theorem \ref{thm:main} is entirely algebraic. 

\item In particular, Conjecture~\ref{conj'} ensures the positivity of the coefficients $\alpha_2,\al_4$, which is rather unexpected in view of \eqref{eq:al2kdefn} and~\eqref{eq:expectation}.   While it does follow from~\eqref{eq:al4} that $\al_4> 0$, most of the effort of this paper will be to show from that formula that in fact $\alpha_4\sim n^2$. 

\item 
Equation \eqref{eq:al2} is almost immediate and shows that $\al_2$ indeed grows linearly in $n$, as required for \eqref{eq:necessary'}. It should be compared with the formula for the $\lam^2$-coefficient for the Almost-Mathieu potential \eqref{eq:AMO}, which remains \emph{bounded} in $n$. See Appendix \ref{app:AMO}.

\item
Equation \eqref{eq:al4} is ultimately a consequence of completing a square appropriately. Its advantages over the original formula \eqref{eq:al2kdefn} for $k=2$ are twofold: (a) The parity conditions have almost completely disappeared, and (b) it features only a \emph{modulus} of Weyl sums, so bounding it from below is more feasible (though still non-trivial). Indeed, a lower bound on the Weyl sums appearing in \eqref{eq:al4} is the content of our other main results.
\e{enumerate}
\e{rmk}

Looking beyond this paper, it seems difficult to deduce that $\alpha_6\sim n^3$. However, it is possible to give lower bounds on the top coefficients which are consistent with Conjecture \ref{conj'}. 

\be{prop}[Top coefficients]\label{prop:top}
There exist constants $c_1,c_2>0$ such that 
\beq
\al_{2n}\geq c_1^n,\qquad \al_{2n-2}\geq c_2^n.
\eeq
\e{prop}

%

\subsection{Lower bounds on quadratic Weyl sums}
Recall that $e[x]=\exp(2\pi i x)$. By Theorem \ref{thm:main}, we have
\beq\label{eq:al4LB}
\al_4\geq 8\sum_{m=1}^{\fl{\frac{n-1}{2}}}\l|\sum_{l=1}^m  \:  e\l[\om\l(l^2-l\r)\r]\r|^2.
\eeq
Considering the conjectured bounds \eqref{eq:necessary'}, the next question is whether we can find irrational numbers $\om\in[0,1]$ and a constant $C>0$ such that we have the bound
\beq\label{eq:nextquestion}
8\sum_{m=1}^{\fl{\frac{n-1}{2}}}\l|\sum_{l=1}^m  \:  e\l[\om\l(l^2-l\r)\r]\r|^2\geq Cn^2+o(n^2),
\eeq
as $n\to\infty$. The two main results presented in this section provide positive answers to this question, taking complementary perspectives. 

The sum 
\begin{align}\label{def:Sm}
S_m(\omega):=\sum_{j=1}^m e[\omega(j^2-j)],
\end{align}
with $\om$ irrational, is an example of an exponential sum. The study of exponential sums has a rich history with close ties to analytic number theory; a classic reference is Montgomery's book \cite{Mont}. Specifically, \eqref{def:Sm} is a \emph{quadratic Weyl sum}, first analyzed by Weyl \cite{W1,W2}. We review methods for estimating Weyl sums in the next section.\\

We come to our \emph{second main result}. For every large $n$, it establishes the existence of a ``good set'' $\Om_n\subset[0,1]$ of uniformly positive Lebesgue measure, such that the lower bound on the Weyl sums holds for all $\om$ in the good set. 

\be{thm}[Second main result]\label{thm:main2}
There exist universal constants $\de,C_0>0$ and an integer $n_0\geq 1$ such that, for every integer $n\geq n_0$, there exists a subset $\Om_n\subset [0,1]$ of Lebesgue measure at least $\de$ such that for all $\om\in\Om_n$, we have
\beq\label{eq:main2}
\sum_{m=1}^{n-1} 
\l|S_m(\om)\r|^2\geq C_0n^2.
\eeq
\e{thm}

Notice that \eqref{eq:main2} is in line with the intuition that Weyl sums should scale in accordance with the central limit theorem, i.e., that $S_m(\om)$ should be of the order $\sqrt{m}$, for irrational $\om$. 

The proof of Theorem \ref{thm:main2} is probabilistic and based on the second moment method. A crucial input are moment asymptotics derived in \cite{JvH} from the central limit theorem for purely quadratic Weyl sums. 
We briefly discuss the main technical difficulties and how we address them in the following section. 

\be{rmk}[The constants]
The constants $C_0$ and $\de$ are semi-explicit; they are given by the formulae
$$
C_0:=C_{JVH}^2\frac{(\sqrt{2}-1)^2 }{64},\qquad \de:=C_{JVH}^2\frac{(\sqrt{2}-1)^2 }{8}.
$$
Here $C_{JVH}:=\int_0^\infty \Phi(x)\d x>0$ is a quantity from \cite{JvH}. It is defined in terms of the $m\to\infty$ limit, $\Phi(x)$, of the distribution function of the Weyl sum $m^{-1/2} \sum_{j=1}^m e[\om j^2]$. (The fact that this limit exists is exactly the central limit theorem, Theorem 3 of \cite{JvH}.) Hence, numerical information on $C_{JVH}$ translates directly into numerical information on $C_0$ and $\de$.
\e{rmk}

The fact that the good sets have uniformly positive measure allows us to obtain, for every subsequence $n_k\to\infty$, a fixed set $\Om\subset[0,1]$, that is good along a subsubsequence.

\be{cor}\label{cor:main2}
Let $\eps>0$ and let $(n_k)_{k\geq 1}$ be a subsequence of the integers. There exists a set $\Om\subset[0,1]$ of Lebesgue measure at least $\de$ such that, for every $\om\in\Om$, we have
$$
\sum_{m=1}^{n_{k_l}-1} 
|S_m(\om)|^2\geq C_0n^2,
$$
along a subsubsequence $n_{k_l}\to\infty$.
\e{cor}

The good set $\Om$ is defined as the $\limsup$ of the good sets $\Om_n$; the corollary then follows from Theorem \ref{thm:main2} by a variant of the converse Borel-Cantelli lemma.

\be{rmk}
\be{enumerate}[label=(\roman*)]
\item We emphasize that the result holds for arbitrary subsequences  $(n_k)_{k\geq 1}$. In particular, one can take $n_k=k$.
\item
Combining the estimates \eqref{eq:al4LB} and \eqref{eq:main2}, we obtain the following lower bound for all $\om\in\Om$:
$$
\al_4\geq C_0n^2,
$$
along the subsubsequence $n_{k_l}\to\infty$. Hence, Corollary \ref{cor:main2} verifies the conjectured bound \eqref{eq:nextquestion}, for all irrational $\om\in\Om$ and along subsequences.
\e{enumerate}
\e{rmk}

Our \emph{third main result} is a complementary result to Corollary \ref{cor:main2}. Concerning $\om$, it is stronger because it holds for Lebesgue-almost every $\om\in[0,1]$. Moreover, it yields a numerically explicit lower bound. Concerning $n$, it is weaker because it only holds along one special subsequence. 

\be{thm}[Third main result]\label{thm:main3}
For Lebesgue almost-every $\omega\in[0,1]$, there exists a subsequence of $n\to\infty$, along which we have
\begin{equation}\label{eq:2}
\sum_{m=1}^n |S_m(\omega)|^2\geq 2 n^2.
\end{equation}
\e{thm}

\be{rmk}
\be{enumerate}[label=(\roman*)]
\item The proof of Theorem \ref{thm:main3} is based on an asymptotic formula of Fiedler-Jurkat-K{\"o}rner \cite{FJK} which says that Weyl sums become ``unusually'' large (i.e., large on the diffusive scale $\sqrt{m}$) along special subsequences determined by the rational approximations to $\om$ (see also \cite{FK}).
\item The constant $2$ on the right-hand side of \eqref{eq:2} is not special and can be replaced by larger numbers.
\item When we apply \eqref{eq:2} to \eqref{eq:al4LB}, we see that, for almost every irrational $\om\in [0,1]$,  the conjectured bound \eqref{eq:nextquestion} holds, with $C=4$, along a certain subsequence.
\e{enumerate}
\e{rmk}

 In summary, together, Theorems \ref{thm:main} and \ref{thm:main2} verify Conjecture \ref{conj'} up to second-order in perturbation theory, taking a different order of quantifiers for $n$ and $\om$.\\

We close this section with a cautionary remark.

\be{rmk}[Rational frequency]
When the frequency $\om=p/q$ is rational, the quadratic Weyl sums $\sum_{j=1}^{m} e\l[\om(j^2-j)\r]$ will typically be of order $m$ (not $\sqrt{m}$). However, in the rational case, the averages over $x,y$ in the definition of $L(\lam,0)$ are not justified by the ergodic theorem and they can lead to a positive Lyapunov exponent because $E=0$ lies in a spectral gap for some values of $x,y$. Therefore, the Lebesgue null set of rational $\om$ should be ignored when interpreting Theorems \ref{thm:main2} and \ref{thm:main3}. 
\e{rmk}

In the next section, we provide some background on the analysis of Weyl sums for the benefit of readers with mathematical physics and spectral theory backgrounds, and we explain how the methods we use fit into the general landscape.

\subsection{Discussion on Weyl sums}
\label{ssect:weyl}
A general quadratic Weyl sum is of the form
$$
S_m(\om,\xi):=\sum_{j=1}^m e[\om j^2+\xi j]
$$
where $\om\in[0,1]\setminus \mathbb Q$ and $\xi\in[0,1]$ are parameters. (Often, the first and last term of the sum are halved so that it is better approximated by an integral.) For irrational $\om$, we expect Weyl sums to live on the ``diffusive scale'' $\sqrt{m}$, indicating the random-like behavior of the quadratic exponentials. (For rational $\om=p/q$, these sums are called ``Gauss sums'' and they are asymptotically much larger, of order $m$, unless $q$ grows with $m$.)

The analysis of Weyl sums has a long history in harmonic analysis, ergodic theory and analytic number theory. Weyl \cite{W1,W2} originally estimated these sums via a method now known as ``Weyl differencing''. Improvements of his approach by van der Corput, Vinogradov and others have become ubiquitous techniques in the study of exponential sums \cite{Mont}.

A classical approach to Weyl sums is due to Hardy and Littlewood \cite{HL}, who viewed Weyl sums as a finitary analog of the Jacobi theta function and established the \emph{approximate functional equation}
\beq\label{eq:HL}
S_m(\om,\xi)=\sqrt{\frac{i}{2\om}} e\l[\frac{-\xi^2}{4\om}\r] S_{\floor{2\om m}}\l(-\frac{1}{4\om},\frac{\xi}{2\om}\r)+O(\om^{-1/2}).
\eeq
Notice that the Weyl sum on the right-hand side has macroscopically fewer terms, $\floor{2\om m}$ compared to $m$ (using that one may assume $\om\in[0,1/2]$ by symmetry and periodicity arguments).  We can iterate this procedure, replacing $-\frac{1}{4\om}$ by its fractional part at every step. With the advent of computers, it was possible to study the curves traced out by the Weyl sums in the complex plane (linearly interpolated). In 1976, Lehmer \cite{Lehmer} observed in this way that incomplete Gauss sums (the case of rational $\om$) form intricate self-similar spiral patterns (``curlicues'') which lie inside a ball of radius proportional to $\sqrt{m}$ (like a random walk would) and whose fine structure depends critically on the \emph{arithmetic properties} of $\om$. (For example, Hardy-Littlewood showed that $S_m\leq O(\sqrt{m})$ if $\om$ is of bounded type. Notice also that the fixed points of the dynamical system $\om\to -\frac{1}{4\om} +\floor{\frac{1}{4\om}}$ appearing in \eqref{eq:HL} are quadratic irrationals.) From the modern perspective, Hardy and Littlewood's formula \eqref{eq:HL} may be seen as a renormalization transformation which groups together curlicues at the smallest scales \cite{BerryGoldberg,CK}. However, it is notoriously difficult to control error terms in this procedure, even given the improved error bounds established later \cite{CK,FK,Mordell,Wilton}. Instead, we will rely on more robust modern variants, as we describe next.\\

Let us now return to our problem at hand --- establishing Theorems \ref{thm:main2} and \ref{thm:main3}. There are three main technical difficulties: (i) we aim for a \emph{lower} bound of appreciable size (which requires good control on the asymptotics); (ii) we need an estimate that holds for a \emph{sum of Weyl sums}, i.e., the estimates on $|S_m|$ need to hold simultaneously for various $m$; (iii) the Weyl sums are not purely quadratic, i.e., they feature the additional linear term $\xi=-\om j$ in the exponential. 

The proof of Theorem \ref{thm:main2} is based on the \emph{second moment method}. The main technical issue (i) is to obtain the correct asymptotics: The higher moment can be computed explicitly by solving a simple Diophantine equation (Lemma \ref{lm:computation}). For the lower moment, we invoke an asymptotic formula due to Jurkat-Van Horne \cite{JvH}, which is a consequence of their central limit theorem for purely quadratic Weyl sums.  The technical difficulty (ii) is easily addressed by linearity of the expectation and Cauchy-Schwarz. Regarding (iii), the key observation is that the sum with the linear term can be rewritten as a purely quadratic Weyl sum over odd integers (see the proof of Lemma \ref{lm:firstmoment}).

Theorem \ref{thm:main3} is instead based on an \emph{asymptotic formula} for Weyl sums established by Fiedler, Jurkat and K{\"o}rner \cite{FJK} which requires $\om$ to be close to a rational number. By a variant of Khinchin's classical result, this  situation occurs infinitely often for Lebesgue almost-every $\om$. This yields a good lower bound, thereby addressing (i), roughly speaking because Weyl sums are very large when $\om=p/q$ is exactly rational. Addressing the technical issue (ii) requires precise estimates on the relevant scales involved. Regarding (iii), the results in \cite{FJK} are fortunately general enough to allow for linear terms.

In closing, we remark that, in recent years, Weyl-sum asymptotics have also been established by using ergodic theory \cite{Cel,CM,Marklof}. As these techniques might become relevant for extending the results in this paper, we briefly discuss them in Appendix B.

\subsection{A reformulation in terms of Diophantine equations}
\label{sect:diophantine}
Finally, we reformulate Conjecture \ref{conj} entirely as a counting problem for Diophantine equations. This is inspired by the cluster expansion method from statistical mechanics, where the convergence radius of a series representation for the logarithm is a posteriori found to be much larger than a naive guess would suggest. 

We call $\beta_{2k}$ the following (non-averaged) analog of $\al_{2k}$. Let $\beta_0=2$ and let
$$
\beta_{2k}:=\sum_{\substack{0\leq k_1,k_2\leq  n\\ k_1+k_2=2k}} (-1)^{\frac{k_1-k_2}{2}} 
\sum_{\substack{1\leq j_1<\ldots<j_{k_1}\leq n\\ 1\leq l_1<\ldots<l_{k_2}\leq n}} \ind_{\clubsuit}(\ul{j},\ul{l})v_{j_1}\ldots v_{j_{k_2}} v_{l_1}\ldots v_{l_{k_2}}
$$
for $k\geq 1$.
By examining the first part of the proof of Proposition \ref{prop:pertseries}, we see that $\beta_{2k}$ are the series coefficients for $\Tr[M_n^*M_n]$ and so
$$
L(\lam,0)=\lim_{n\to\infty}\frac{1}{n}\mathbb E_{\T^2}\log \Tr[M_n^*M_n]
=\lim_{n\to\infty}\frac{1}{n}\mathbb E_{\T^2}\l[\log\sum_{k=0}^n \beta_{2k}\lam^{2k}\r].
$$
Let us now also take the average over $\om\in [0,1]$; denote the total average (over $x,y,\om$) by $\mathbb E_{\T^3}$. If we can show that the resulting quantity is strictly positive, than there must exist a ``good set'' $\Om\subset[0,1]$ of positive Lebesgue measure such that $L(\lam,0)>0$ --- confirming Conjecture \ref{conj} for all $\om\in\Om$.

For fixed $n$, we may expand the logarithm as a power series in $\lam$ and then take the expectation $\mathbb E_{\T^3}$. Note that this expansion is a priori only formal for sufficiently large $n$, since the averages of the coefficients $\beta_{2k}$ grow with $n$. Nonetheless, Conjecture \ref{conj} can be reformulated as saying that, at every order in $\lam$, the averaged coefficients cancel precisely to yield a quantity of order $n$ (which is then multiplied by the prefactor $\frac{1}{n}$ from above). 

If the cancellations occur at every order to yield a quantity of order $n$, then this shows that the series expansion for the logarithm converges for small enough $\lam$ and arbitrarily large $n$. (As we mentioned before, this phenomenon occurs frequently in statistical mechanics where the cluster expansion method can be used to calculate the ``partition function'', a quantity which is formally similar to the Lyapunov exponent.)\\

For example, the $\lam^4$ term of the logarithm is 
$$
\frac{1}{2}\l(\mathbb E_{\T^3}[\beta_4]-\frac{\mathbb E_{\T^3}[\beta_2^2]}{4}\r).
$$
(The additional $1/2$ factors enter because $\beta_0=2$.) Note that $\mathbb E_{\T^3}[\beta_4]=\int_0^1 \al_4\d\om$ counts the number of solutions to Diophantine equations and can be seen to be at least of order $n^2$. This has to be canceled rather precisely by the other term $\frac{\mathbb E_{\T^3}[\beta_2^2]}{4}$, which also counts solutions to other Diophantine equations, in order to obtain an order $n$ quantity. 

The analogous statements at every order in $\lam^2$ provide a reformulation of Conjecture \ref{conj} as a counting problem for solutions to Diophantine equations, though, admittedly, a rather non-trivial one.

%

\section{Derivation of the perturbation series and Proposition \ref{prop:top}}
\subsection{Proof of Proposition \ref{prop:pertseries}}
For $E=0$, we may write each transfer matrix $A_j$ as 
$$
A_j=\ttmatrix{0}{-1}{1}{0}-\lam v_j \ttmatrix{1}{0}{0}{0}.
$$
We expand the expression $\Tr[M_n^* M_n]$ as a polynomial in $\lam$. A straightforward computation using cyclicity of the trace then shows that
$$
\curly{P}_n(\lam,E)=\mathbb E_{\mathbb T^2} \l[\Tr[M_n^* M_n]\r]=\sum_{k=0}^{n} \al_{2k} \lam^{2k},
$$
with coefficients given by
$$
\begin{aligned}
&\al_{2k}=\\
&\sum_{\substack{0\leq k_1,k_2\leq  n\\ k_1+k_2=2k}} (-1)^{\frac{k_1-k_2}{2}}
\sum_{\substack{1\leq j_1<\ldots<j_{k_1}\leq n\\ 1\leq l_1<\ldots<l_{k_2}\leq n}} \ind_{\clubsuit}(\ul{j},\ul{l})\mathbb  E_{\mathbb T^2}[ v_{j_1}\ldots v_{j_{k_1}} v_{l_1}\ldots v_{l_{k_2}}].
\end{aligned}
$$
We remark that this formula holds for any choice of the potential $v_j$. 

To prove Proposition \ref{prop:pertseries}, it remains to show that only terms with $k_1-k_2\equiv 0\mod 4$ contribute to the sum. Notice that this condition is equivalent to $\frac{k_1-k_2}{2}\equiv 0\mod 2$ and so $(-1)^{\frac{k_1-k_2}{2}}=1$. 

This part uses the skew-shift structure. We consider
$$
\begin{aligned}
\sum_{\substack{1\leq j_1<\ldots<j_{k_1}\leq n\\ 1\leq l_1<\ldots<l_{k_2}\leq n}} \ind_{\clubsuit}(\ul{j},\ul{l}) \mathbb E_{\mathbb T^2}[ v_{j_1}\ldots v_{j_{k_1}} v_{l_1}\ldots v_{l_{k_2}}].
\end{aligned}
$$
The expectation is given by equation \eqref{eq:expectation}. From it, we see that a non-zero contribution can only come from pairs of vectors $(\ul{j},\ul{l})$ for which there exists a vector $\ul{a}\in \{\pm 1\}^{2k}$ such that $\ul{a}\perp (\ul{j},\ul{l})$, i.e., 
$$
\sum_{s=1}^{k_1} a_s j_s+\sum_{s={k_1+1}}^{2k} a_s l_{s-k_1}=0. 
$$
Since $a_s\in \{\pm 1\}$, this linear relation implies that among the entries of $\ul{j}$ and $\ul{k}$, we must have \emph{an even number of odd entries}. That is, if we define
$$
\begin{aligned}
\nu(\ul{j}):=&|\setof{1\leq s\leq k_1}{j_s \textnormal{ odd}}|
\end{aligned}
$$
(with $|\cdot|$ denoting the cardinality of a set) and the analogous quantity $\nu(\ul{l})$, then we have
\beq\label{eq:paritycondition}
\begin{aligned}
\nu(\ul{j})+\nu(\ul{l})\equiv 0\mod 2.
\end{aligned}
\eeq

 Next, we show that \eqref{eq:paritycondition} implies $\frac{k_1-k_2}{2}\equiv 0\mod 2$. By the $\clubsuit$ condition \eqref{eq:clubdefn}, the parity of each entry of $\ul{j}$ and of $\ul{l}$ alternates. This may be formalized as follows.
$$
\begin{aligned}
\nu(\ul{j})=
\be{cases}
\floor*{\frac{k_1}{2}},\qquad &\textnormal{if $j_1$ is even},\\
\ceil*{\frac{k_1}{2}},\qquad &\textnormal{if $j_1$ is odd}, 
\e{cases}
\end{aligned}
$$
and an analogous formula holds for $\nu(\ul{l})$ with $k_1$ replaced by $k_2$. Suppose that $j_1$ is even; notice that this implies that $l_1$ is even as well. Then \eqref{eq:paritycondition} yields
$$
\floor*{\frac{k_1}{2}}+\floor*{\frac{k_2}{2}}\equiv 0\mod 2.
$$
By subtracting $k_2$ from this equation, and distinguishing cases according to the parity of $k_2$, we conclude that
$$
\frac{k_1-k_2}{2} \equiv 0\mod 2
$$
as claimed. A similar argument holds if $j_1$ and $l_1$ are odd. This finishes the proof of Proposition \ref{prop:pertseries}.
\qed

\subsection{Proof of Proposition \ref{prop:top}}
We first consider $\al_{2n}$. The only contributions to \eqref{eq:al2kdefn} come from the ``diagonal'' $k_1=k_2=n$ and $\ul{j}=\ul{l}=(1,2,\ldots,n)$. Hence, by Jensen's inequality,
$$
\al_{2n}=\mathbb E_{\T^2}\l[\prod_{j=1}^n v_j^2\r]
\geq \exp \l(\sum_{j=1}^n\mathbb E_{\T^2}\l[ \log (v_j^2)\r]\r)
=c_1^n,
$$
with $c_1:=\exp \int_0^1 \log \cos^2(2\pi x)\d x>0$.

For $\al_{2n-2}$, the only contributions to \eqref{eq:al2kdefn} come from $k_1=k_2=n-1$ and either $\ul{j}=\ul{l}=(1,2,\ldots,n-1)$
or $\ul{j}=\ul{l}=(2,\ldots,n)$. An analogous application of Jensen's inequality finishes the proof of Proposition \ref{prop:top}.
\qed

\section{Proof of the identities in Theorem \ref{thm:main}}
\subsection{Proof of formula \eqref{eq:al2} for $\al_2$}
By Proposition \ref{prop:pertseries}, we have
$$
\begin{aligned}
\al_2=\sum_{1\leq j_1,j_2\leq n} \ind_{(e,e)}(j_1,j_2)\mathbb E_{\mathbb T^2} [v_{j_1}v_{j_2}] +p.c.
\end{aligned}
$$
Here and in the following, we write $p.c.$ for the ``parity conjugate'' of the preceding expression, i.e., the same expression with all appearances of $\ev$ and $\od$ interchanged. 

We compute the expectation via \eqref{eq:expectation}. Since $a_1=1$, we must have $a_2=-1$ and so
$$
\ind_{\ul{a}\perp \ul{j}}=\ind_{j_1=j_2}.
$$
This implies $a_1 j_1^2+a_2j_2^2=0$ and therefore $\mathbb E_{\mathbb T^2} [v_{j_1}v_{j_2}]=2\ind_{j_1=j_2}$ in the formula for $\al_2$ above. We see that the sum over $j_2$ collapses and \eqref{eq:al2} is proved.
\qed

\subsection{Proof of formula \eqref{eq:al4} for $\al_4$}
For this part, it is convenient to introduce some notation. Recall that we denote the sets of even (odd) integers by $\ev$ and $\od$, respectively. Given a choice of parities $\pi_1,\ldots,\pi_k\in \{\ev,\od\}$ and real numbers $r_1,\ldots,r_k$, we define  
$$
[\pi_1^{r_1}  \ldots \pi_k^{r_k}]:=\sum_{1\leq j_1<\ldots<j_k\leq n} \ind_{(\pi_1,\ldots,\pi_k)}(j_1,\ldots,j_k)  v^{r_1}_{j_1}\ldots v^{r_k}_{j_k}.
$$

\medskip

We begin by applying Proposition \ref{prop:pertseries} to find
\beq
\label{eq:rewriteal2}
\al_4=\mathbb E_{\mathbb T^2}\l[2[\ev\od\ev\od]+[\ev\od][\ev\od]+p.c.\r].
\eeq

We may compute the product in \eqref{eq:rewriteal2}. Namely
$$
\begin{aligned}
[\ev\od][\ev\od]
=2[\ev\od\ev\od]+4[\ev\ev\od\od]+2[\ev\ev\od^2]+2[\ev^2\od\od]+[\ev^2\od^2].
\end{aligned}
$$
The following lemma provides the expectation of all terms containing higher powers of $v_j$.

\be{lm}\label{lm:31term}
We have
\begin{align}
\label{eq:lm1}
\mathbb E_{\mathbb T^2}\l[[\ev\ev\od^2]+[\ev^2\od\od]+p.c.\r]=&0,\\
\label{eq:lm2}
\mathbb E_{\mathbb T^2}[\ev^2\od^2+p.c.]=&n^2-\ind_\od(n).
\end{align}
\e{lm}

We postpone the proof of this lemma for now. Upon returning to \eqref{eq:rewriteal2}, Lemma \ref{lm:31term} implies that
$$
\al_4=4\mathbb E_{\mathbb T^2} \l[[\ev\od\ev\od]+[\ev\ev\od\od]\r]+p.c.+n^2-\ind_\od(n).
$$
Recalling our notation, we are led to consider
\beq\label{eq:ledto}
\begin{aligned}
&4\mathbb E_{\mathbb T^2} \l[[\ev\od\ev\od]+[\ev\ev\od\od]\r]\\
=&4\sum_{1\leq j_1<j_2<j_3<j_{4}\leq n}  \ind_{(\ev,\od)}(j_1,j_4)\l(\ind_{(\ev,\od)}+\ind_{(\od,\ev)}\r)(j_2,j_3) 
\mathbb E_{\mathbb T^2} [v_{j_1}\ldots v_{j_4}]
\end{aligned}
\eeq
(plus its parity conjugate). We now compute the expectation via \eqref{eq:expectation}. There are three choices  of $\ul{a}\in\{\pm 1\}^4$ such that $a_1=1$ and $\ul{a}\perp \ul{1}$; see the table.

\be{table}[h]
\be{center}
\be{tabular}{c|c|c|c|c|}
 & $a_1$ & $a_2$ & $a_3$&	$a_4$\\
 \hline
 $(I)$ & $1$ & $1$ & $-1$&	$-1$\\
 $(II)$ & $1$ & $-1$ & $1$&	$-1$\\
 $(III)$ & $1$ & $-1$ & $-1$&	$1$
\e{tabular}
\e{center}
\caption{Different choices of vectors $\ul{a}\in \{\pm 1\}^4$ with $a_1=1$ and $\ul{a}\perp \ul{1}$.}\label{table:achoices}
\e{table} 

 Considering the fact that $j_1<j_2<j_3<j_4$, we find that only case (III) yields a non-zero indicator function $\ind_{\ul{a}\perp \ul{j}}$, namely
$$
\ind_{(1,-1,-1,1)\perp \ul{j}}=\ind_{j_1+j_4=j_2+j_3}.
$$
An important observation is that the parity conditions in \eqref{eq:ledto} amount precisely to specifying that $j_1+j_4=j_2+j_3$ is odd, i.e.,
$$
\begin{aligned}
&4\mathbb E_{\mathbb T^2} \l[[\ev\od\ev\od]+[\ev\ev\od\od]\r]+p.c.\\
=&8\sum_{s=3}^{2n-1} \ind_\od(s) \sum_{1\leq j_1<j_2<j_3<j_{4}\leq n}  \ind_{j_1+j_4=j_2+j_3=s}\, e\l[\frac{\om}{2}\l(j_1^2+j_4^2-j_2^2-j_3^2\r)\r].
\end{aligned}
$$
Next we observe that, conditional upon $j_1+j_4=j_2+j_3$, the condition $j_1<j_2<j_3<j_{4}$ is equivalent to $j_1<j_2<j_3$ and $j_1<j_4$ (where the last constraint is in fact redundant). We can use this fact to complete a square in the above expression:
$$
\begin{aligned}
&8\sum_{s=3}^{2n-1} \ind_\od(s) \sum_{1\leq j_1<j_4\leq n} \ind_{j_1+j_4=j_2+j_3=s}\,\: e\l[\frac{\om}{2}\l(j_1^2+j_4^2\r)\r]\\
&\times\sum_{\substack{1\leq j_2<j_3\leq n\\ j_2>j_1}} \ind_{j_2+j_3=s} \: e\l[\frac{\om}{2}\l(-j_2^2-j_3^2\r)\r]\\
=&4\sum_{s=3}^{2n-1} \ind_\od(s) 
\l(\l|\sum_{1\leq j_1<j_4\leq n} \ind_{j_1+j_4=s} \:  e\l[\frac{\om}{2}\l(j_1^2+j_4^2\r)\r]\r|^2
-\sum_{1\leq j_1<j_4\leq n} \ind_{j_1+j_4=s} \r)\\
=&4\sum_{s=3}^{2n-1} \ind_\od(s) 
\l(\l|\sum_{j_1=1}^{n-1}\ind_{j_1<s-j_1\leq n} \:  e\l[\om\l(j_1^2-s j_1\r)\r]\r|^2
-\sum_{j_1=1}^{n-1} \ind_{j_1<s-j_1\leq n} \r)\\
=&4\sum_{m=1}^{n-1} 
\l(\l|\sum_{j_1=1}^{m}\ind_{j_1\geq 2m+1-n} \:  e\l[\om\l(j_1^2-(2m+1) j_1\r)\r]\r|^2
-\sum_{j_1=1}^{m}\ind_{j_1\geq 2m+1-n} \r).
\end{aligned}
$$
Now we change the inner summation variable to $l:=m+1-j_1$ and obtain
$$
\begin{aligned}
&4\sum_{m=1}^{n-1} 
\l(\l|\sum_{l=1}^{\min\{m,n-m\}}  \:  e\l[\om\l(l^2-l\r)\r]\r|^2
-\sum_{l=1}^{\min\{m,n-m\}} 1\r)\\
=&4\sum_{m=1}^{\fl{\frac{n}{2}}} 
\l|\sum_{l=1}^m  \:  e\l[\om\l(l^2-l\r)\r]\r|^2
+4\sum_{m=1}^{\fl{\frac{n-1}{2}}}\l|\sum_{l=1}^m  \:  e\l[\om\l(l^2-l\r)\r]\r|^2\\
&-n^2+\ind_\od(n).
\end{aligned}
$$

This finishes the proof of Theorem \ref{thm:main}, modulo Lemma \ref{lm:31term}.

\subsection{Proof of Lemma \ref{lm:31term}}
For \eqref{eq:lm1}, we consider
$$
\begin{aligned}
\mathbb E_{\mathbb T^2}[\ev\ev\od^2]
=2\sum_{1\leq j_1<j_2<j_3=j_4\leq n} \ind_{(\ev,\ev,\od)}(j_1,j_2,j_3) \sum_{\substack{\ul{a}\in\{\pm 1\}^4\\ a_1=1\\ \ul{a}\perp \ul{1}, \ul{j}}} c\l[\frac{\om}{2} \sum_{s=1}^{4} a_s j_s^2\r].
\end{aligned}
$$
We recall Table \ref{table:achoices}. One may check that $\ind_{\ul{a}\perp \ul{j}}=0$ in each case (I)-(III) separately, and so $\mathbb E_{\mathbb T^2}[\ev\ev\od^2]=0$. Since the argument holds independently of parity, it also gives $\mathbb E_{\mathbb T^2}[\ev^2\od\od]=0$ and hence \eqref{eq:lm1}.\\

It remains to prove \eqref{eq:lm2}. We consider
$$
\mathbb E_{\mathbb T^2}[\ev^2\od^2]=2\sum_{1\leq j_1=j_2<j_3=j_4\leq n} \ind_{(\ev,\od)}(j_1,j_3) \sum_{\substack{\ul{a}\in\{\pm 1\}^4\\ a_1=1\\ \ul{a}\perp \ul{1}, \ul{j}}} c\l[\frac{\om}{2} \sum_{s=1}^{4} a_s j_s^2\r].
$$
Recall Table \ref{table:achoices} once more. For case (I), we obtain
$$
\ind_{(1,1,-1,-1)\perp \ul{j}}=\ind_{2j_1= 2j_3}=0,
$$
because $j_1<j_3$. For cases (II) and (III), we obtain a non-zero contribution. Indeed
$$
\ind_{(1,-1,1,-1)\perp \ul{j}}=\ind_{(1,-1,-1,1)\perp \ul{j}}=\ind_{0=0}=1.
$$
Since $\sum_{s=1}^{4} a_s j_s^2=0$ in cases (II) and (III), we find
$$
\begin{aligned}
\mathbb E_{\mathbb T^2}[\ev^2\od^2+p.c.]
=&4\sum_{1\leq j_1=j_2<j_3=j_4\leq n} \l(\ind_{(\ev,\od)}(j_1,j_3)+\ind_{(\od,\ev)}(j_1,j_3)\r)\\
=&4 \sum_{\substack{1\leq j\leq n\\ j\in o}} \sum_{\substack{1\leq l\leq n\\ l\in e}} 1\\
=&n^2-\ind_\od(n).
\end{aligned}
$$
This finishes the proof of Lemma \ref{lm:31term} and hence of Theorem \ref{thm:main}.
\qed

\section{Proof of the probabilistic lower bound (Theorem \ref{thm:main2})}

\subsection{The second moment method}
The proof is based on the second moment method, i.e.,

\be{prop}[Paley-Zygmund inequality]\label{prop:PZ}
Let $Z\geq 0$ be a random variable and let $\theta\in[0,1]$. Then, it holds that
\beq\label{eq:PZ}
\mathbb P(Z>\theta \E [Z])\geq (1-\theta)^2\frac{(\E[Z])^2}{\E[Z^2]}.
\eeq
\e{prop}

\be{proof}
By Cauchy-Schwarz
$$
\E[Z]=\E[Z\ind_{Z\leq \theta \E[Z]}]+\E[Z\ind_{Z>\theta \E[Z]}]
\leq \theta \E[Z]+\sqrt{\E[Z^2]P(Z>\theta \E[Z])},
$$
and \eqref{eq:PZ} follows by rearranging.
\e{proof}

We recall that $S_{m}= \sum_{j=1}^{m} e\l[\om (j^2-j)\r]$. We will apply the Paley-Zygmund inequality to the family of random variables
\beq\label{eq:Zndefn}
Z_n(\om):=\sqrt{\sum_{m=1}^{n-1} |S_m(\om)|^2},
\eeq
which are obtained by sampling the frequency $\om\in [0,1]$ at random, according to uniform (i.e., Lebesgue) measure. We write $\mathbb E$ for expectation with respect to that measure. 

The following two lemmas allow us to control the first and second moment of $Z_N$, so that we can use \eqref{eq:PZ}.

\be{lm}\label{lm:computation}
For any integer $n\geq 1$, we have
$$
\E[Z_n^2]=\frac{n(n-1)}{2}.
$$
\e{lm}

\be{lm}\label{lm:firstmoment}
There exists a constant $C_1>0$ such that 
$$
\liminf_{n\to\infty} \frac{\E[Z_n]}{n}\geq C_1.
$$
\e{lm}

Lemma \ref{lm:computation} is a straightforward computation; at its core stands the solution of a simple Diophantine equation.
Lemma \ref{lm:firstmoment}, on the other hand, requires as an input the asymptotics of the first moments of purely quadratic Weyl sums (without a linear term), which are a consequence of the central limit theorem of Jurkat and Van Horne \cite{JvH}.  

Before, we prove Lemmas \ref{lm:computation} and \ref{lm:firstmoment}, let us see that they imply Theorem \ref{thm:main2}. 

\be{proof}[Proof of Theorem \ref{thm:main2}]
Let $\theta=1/2$. We combine Proposition \ref{prop:PZ} with Lemmas \ref{lm:computation} and \ref{lm:firstmoment} to obtain
$$
\begin{aligned}
\liminf_{n\to\infty}\mathbb P(Z_n> \E[Z_n]/2)
\geq& \frac{1}{4}\liminf_{n\to\infty}\frac{(\E[Z_n])^2}{\E[Z_n^2]}\geq \frac{C_1^2}{2}.
\end{aligned}
$$
Moreover, by Lemma \ref{lm:firstmoment}, we have $\E[Z_n]/2\geq n C_1/4$ for all large enough $n$. Hence, we have shown that
$$
\liminf_{n\to\infty}\mathbb P\l(\frac{1}{n^2}\sum_{m=1}^{n-1} |S_m(\om)|^2> \frac{C_1^2}{16}\r)
=\liminf_{n\to\infty}\mathbb P\l(Z_n> \frac{C_1}{4}n\r)\geq \frac{C_1^2}{2}.
$$
We can now define the ``good'' sets as
\beq\label{eq:Gndefn}
\Om_{n}:=\setof{\om\in [0,1]}{\frac{1}{n^2}\sum_{m=1}^{n-1} |S_m(\om)|^2> \frac{C_1^2}{16}}.
\eeq
The statement above shows that $\Om_n$ has uniformly positive Lebesgue measure, with the lower bound $\de:=\frac{C_1^2}{2}$, for every sufficiently large $n$. This proves Theorem \ref{thm:main2}.
\e{proof}

\subsection{Proof of Lemmas \ref{lm:computation} and \ref{lm:firstmoment}}
\be{proof}[Proof of Lemma \ref{lm:computation}]
By orthonormality of $\{e[j\cdot]\}_{j\in \Z}$ and the fact that $j^2-j=k^2-k$ is equivalent to $j=k$ for positive integers, we have
$$
\begin{aligned}
\mathbb E\l[\sum_{m=1}^{n-1} 
|S_m|^2\r]
:=&\mathbb E\l[\sum_{m=1}^{n-1} 
\l|\sum_{j=1}^{m} e\l[\om(j^2-j)\r]\r|^2\r]\\
=&\sum_{m=1}^{n-1} \sum_{j,k=1}^m \mathbb E
\l[e\l[\om(j^2-j-k^2+k)\r]\r]\\
=&\sum_{m=1}^{n-1} m = \frac{n(n-1)}{2}.
\end{aligned}
$$
This proves Lemma \ref{lm:computation}.
\e{proof}

The proof of Lemma \ref{lm:firstmoment} uses the following result from \cite{JvH}. We define the purely quadratic Weyl sum
$$
W_m(\om):=\sum_{j=1}^{m} e[\om j^2].
$$
We write $f(m)\sim g(m)$, for $\lim_{m\to\infty}\frac{f(m)}{g(m)}=1$.

\be{thm}[\cite{JvH}, Theorem 4]\label{thm:JvH}
There exists a constant $C_{JVH}>0$, such that, as $m\to\infty$,
\beq\label{eq:JvH}
\mathbb E[|W_m|]\sim C_{JVH} m^{1/2}.
\eeq
\e{thm}

We point out that \cite{JvH} use a different convention for the quadratic Weyl sums, where the first terms are halved (though this does not influence the asymptotics) and $\om$ ranges only over $[0,1/2]$. The statement \eqref{eq:JvH} follows from their Theorem 4 by $\Z$-periodicity of $e[\cdot]$ and the reflection symmetry $|W_m(\om)|=|W_m(-\om)|$. We have $C_{JVH}:=\int_0^\infty \Phi(x)\d x>0$ with $\Phi$ the limiting distribution function from Theorem 3 in \cite{JVH}. 

We are now ready to give the

\be{proof}[Proof of Lemma \ref{lm:firstmoment}]
By Cauchy-Schwarz, we have
\beq\label{eq:ZnCS}
Z_n=\sqrt{\sum_{m=1}^{n-1} |S_m|^2}\geq \frac{1}{\sqrt{n-1}}\sum_{m=1}^{n-1} |S_m|.
\eeq
We will bound the right-hand side from below via Theorem \ref{thm:JvH}. The difference between our Weyl sums $S_m$ and the purely quadratic ones treated by \cite{JvH} is the linear term $-j \om$ in the exponential. The key observation is that we may nonetheless reduce the computation to the case of $W_m$ by parity arguments. We decompose
$$
W_m=W_m^\ev+W_m^\od,
$$
where the $\ev$ ($\od$) terms are given by restricting $j$ to the set of even (odd) integers, respectively. 

We have
$$
|S_m(\om)|=\l|\sum_{j=1}^m e[\om(j^2-j)]\r|
=\l|\sum_{j=1}^m e\l[\frac{\om}{4}(2j-1)^2\r]\r|
=\l|W_{2m-1}^\od\l(\frac{\om}{4}\r)\r|.
$$
By the $\Z$-periodicity of $e[\cdot]$, we have $S_m(\om)=S_m(\om+1)$, and therefore, by a change of variable,
$$
\E [|S_m|]
= \E [|S_m(4\cdot)|]
=\E \l[\l|W_{2m-1}^\od\r|\r].
$$
By the triangle inequality and a change of summation index, we obtain the lower bound
$$
\begin{aligned}
\E \l[\l|W_{2m-1}^\od\r|\r]
\geq& \E \l[\l|W_{2m-1}\r|\r]-\E \l[\l|W_{2m-2}^\ev\r|\r]\\
=&\E \l[\l|W_{2m-1}\r|\r]-\E \l[\l|W_{m-1}(4\cdot)\r|\r]\\
=&\E \l[\l|W_{2m-1}\r|\r]-\E \l[\l|W_{m-1}\r|\r].
\end{aligned}
$$
In the last step, we used the periodicity $W_{m-1}(\om)=W_{m-1}(\om+1)$. Now we can apply Theorem \ref{thm:JvH} to conclude that
$$
\liminf_{m\to\infty}\frac{\E [|S_m|]}{\sqrt{m}}
\geq C_{JVH}\liminf_{m\to\infty}\frac{\sqrt{2m-1}-\sqrt{m-1}}{\sqrt{m}}
=(\sqrt{2}-1)C_{JVH}.
$$
Let us fix $\eps\in (0,1)$.
We now apply this estimate to \eqref{eq:ZnCS} and obtain
$$
\begin{aligned}
\liminf_{n\to\infty}\frac{\E [Z_n]}{n}
\geq& \liminf_{n\to\infty}\frac{1}{n\sqrt{n-1}}\sum_{m=\eps n}^{n-1} \E[|S_m(\om)|]\\
\geq& (\sqrt{2}-1)C_{JVH}
\liminf_{n\to\infty}n^{-3/2}\sum_{m=\eps n}^{n} \sqrt{m}\\
\geq& \frac{\sqrt{2}-1}{2}C_{JVH}.
\end{aligned}
$$
The last estimate holds for sufficiently small $\eps$.
This proves Lemma \ref{lm:firstmoment} (with $C_1:=\frac{\sqrt{2}-1}{2}C_{JVH}$) and hence finishes the proof of Theorem \ref{thm:main2}.
\e{proof}

\subsection{Proof of Corollary \ref{cor:main2}}
Recall \eqref{eq:Gndefn} from above. We need to ensure that the event that $\Om_{n_k}$ occurs infinitely often has positive probability. Formally, this event is defined as
$$
\Om:=\limsup_{k\to\infty} \Om_{n_k} :=\bigcap_{K=1}^\infty \bigcup_{k=K}^\infty \Om_{n_k}.
$$
The claim of the corollary can then be compactly written as
\beq\label{eq:BC}
\mathbb P(\Om)\geq \de>0.
\eeq

By Theorem \ref{thm:main2}, we have that $\liminf_{n\to\infty}\mathbb P(\Om_{n})\geq \de>0$.
Recall that any probability measure is continuous on monotone sequences of events. Hence
$$
\begin{aligned}
\mathbb P(\Om)
=&\mathbb P\l(\bigcap_{K=1}^\infty \bigcup_{k=K}^\infty \Om_{n_k}\r)
=\lim_{K\to\infty} \mathbb P\l(\bigcup_{k=K}^\infty \Om_{n_k}\r)\\
\geq& \liminf_{K\to\infty} \mathbb P\l( \Om_{n_K}\r)
\geq \de.
\end{aligned}
$$
This proves \eqref{eq:BC} and hence Corollary \ref{cor:main2}.
\qed

\section{Proof of the almost-sure lower bound (Theorem \ref{thm:main3})}

\subsection{Preliminaries}
The following results were proved in \cite{FJK}. Let $p$ and $q$ be integers and let
$$F(y):=\frac{1}{\sqrt{i}}\int_0^y e^{\pi i t^2}\, dt$$
be the Fresnel integral.

\begin{thm}\cite[Theorem 5]{FJK}\label{thm5:FJK}
Let $\theta$ be a real number, $m$ be a positive integer and $0<\varepsilon\leq 1/2$.
Choose $A$ such that $pq+2A$ is an even integer and that $\theta=A+a$ holds with $|a|\leq 1/2$.
Then, for real $\xi\neq 0$ with $|m\xi+a|\leq 1-\varepsilon$, we have
\beq
\begin{aligned}\label{def:SNFJK}
S_m(p,q,\xi,\theta):=&\sum_{n=1}^m \exp\left(\pi i \left(n^2 \frac{p+\xi}{q}+n\frac{2\theta}{q}\right)\right)\\
=&T_m+O_{\varepsilon}(\sqrt{q}(1+|\xi|q)),
\end{aligned}
\eeq
where
$$T_m=\frac{h}{\sqrt{\xi}}\exp\left(-\pi i \frac{a^2}{q\xi}\right) \Big[F\left(\frac{m\xi+a}{\sqrt{q\xi}}\right)-F\left(\frac{a}{\sqrt{q\xi}}\right)\Big],$$
in which $h$ is a complex number with $|h|=1$.
\end{thm}
\begin{cor}\cite[Corollary of Theorem 5]{FJK}\label{cor5:FJK}
Let $0<|\xi|\leq 1/(4m)$, $0<q\leq 4m$, $a'=a\, \mathrm{sign}\, \xi$, $|a'|\leq 1/2$, $(m^2|\xi|+2ma')/q=2k+\gamma$ for some integer $k$ and $|\gamma|\leq 1$. Then, for some absolute constants $c_1, c_2>0$, we have
\begin{align}
T_m\sim_{c_1,c_2}
\begin{cases}
\frac{m}{\sqrt{q}+m\sqrt{|\xi|}},\ \ \text{if}\ \ -m|\xi|-\sqrt{q|\xi|}\leq a'\leq \sqrt{q|\xi|},\\
\\
\sqrt{q}\left(\frac{m|\xi|}{a'(m|\xi|+a')}+\frac{|\gamma|}{\sqrt{a'(m|\xi|+a')}}\right),\ \ \text{otherwise}.
\end{cases}
\end{align}
where $X\sim_{c_1,c_2}Y$ means $c_1Y\leq X\leq c_2Y$.
\end{cor}

We will also use the following lemma, which is a variant of Khinchin's theorem. 
\begin{lm}\label{lm:Khinchin}
Let $(\psi(n))_{n=1}^\infty$ be a non-increasing sequence of positive numbers such that $\sum_{n=1}^{\infty}\psi(n)$ diverges. Then for Lebesgue almost-every $\omega\in \mathbb{R}$, the inequality 
$$|q\omega-p|<\psi(q)$$
has infinitely many solutions $p,q$ with $2|p$ and $(p,q)=1$.
\end{lm}

We postpone the proof of this lemma to the end of this section. 

\subsection{Proof of Theorem \ref{thm:main3}}
We let $C>0$ be an absolute constant which is chosen at the end of the proof; see \eqref{def:constantC}.
Lemma \ref{lm:Khinchin} with $\psi(n)=(C n)^{-1}$ implies that for Lebesgue almost-every $\omega\in \R$ ,
there are infinitely many solutions $p,q$ with $2|p$ and $(p,q)=1$, and 
\begin{align}\label{eq:pq}
|q (2\omega)-p|<\frac{1}{C q}.
\end{align}
We fix such an $\om\in[0,1]$. Let us label the corresponding sequence of solutions by $p_k, q_k$ with $q_k$ in increasing order. We
set 
$$
\xi_k:=2q_k\omega-p_k,\qquad \textnormal{i.e.,}\quad 2\omega=\frac{p_k+\xi_k}{q_k},
$$
and 
$$
\theta_k:=-q_k \omega=-\frac{p_k}{2}-\frac{\xi_k}{2}.$$ 
With these choices, we have
$$
S_m(\om)=S_m(p_k,q_k,\xi_k,q_k).
$$
We check the conditions of Theorem \ref{thm5:FJK}. By \eqref{eq:pq}, we have
\begin{align}\label{eq:xik}
|\xi_k|<\frac{1}{C q_k}\leq \frac{1}{16 q_k},
\end{align}
provided $C\geq 16$.
Since $p_k$ is an even number, we can take 
\begin{align}\label{def:Akalphak}
A_k=-\frac{p_k}{2},\ \ \text{and}\ \ a_k=-\frac{\xi_k}{2}.
\end{align}
We define the ($\om$-dependent) subsequence $N_k$ by 
$$N_k:=\lfloor \sqrt{C} q_k \rfloor,
$$
and we let $m$ be an integer with $\frac{N_k}{2}\leq m\leq N_k$. We check that 
\begin{align}\label{eq:Nkxik}
m |\xi_k|\leq N_k |\xi_k|\leq \frac{1}{\sqrt{C}}\leq \frac{1}{4}.
\end{align}
and hence in view of \eqref{eq:xik}
$$|m\xi_k+a_k|\leq \frac{1}{4}+\left|\frac{\xi_k}{2}\right|<\frac{1}{2}.$$
Thus Theorem \ref{thm5:FJK} implies that, for some absolute constant $c_3>0$,
\begin{align}\label{eq:SNk}
|S_{m}(\omega)|=|S_{m}(p_k, q_k, \xi_k, \theta_k)|\geq T_m-c_3\sqrt{q_k},
\end{align}
where we used \eqref{eq:xik} to simplify the error term.

Next, we use the first estimate in Corollary \ref{cor5:FJK} to estimate $T_m$. 
In view of \eqref{def:Akalphak} and \eqref{eq:Nkxik}, it remains to compute that
$$q_k\leq 2\lfloor \sqrt{C} q_k \rfloor = 2N_k\leq 4m,$$
and 
$$-m|\xi_k|-\sqrt{q_k|\xi_k|}\leq \frac{|\xi_k|}{2}\leq \sqrt{q_k |\xi_k|}.$$
Hence Corollary \ref{cor5:FJK} implies that there exists an absolute constant $c_1>0$ such that 
\begin{align}\label{eq:Tm}
T_m\geq c_1 \frac{m}{\sqrt{q_k}+m\sqrt{|\xi_k|}}\geq \frac{c_1}{2} \frac{m}{\sqrt{q_k}},
\end{align}
where we used \eqref{eq:xik} in the last inequality.
Combining \eqref{eq:Tm} with \eqref{eq:SNk}, we have
\begin{align}\label{eq:Smlbb}
|S_{m}(\omega)| \geq \frac{c_1}{2} \frac{m}{\sqrt{q_k}} - c_3 \sqrt{q_k}\geq \frac{c_1}{4} \frac{m}{\sqrt{q_k}}\geq \frac{c_1 \sqrt{C}}{4\sqrt{2}}\sqrt{m},
\end{align}
where we used 
$$m\geq \frac{\sqrt{C}}{2}q_k \geq \frac{4c_1}{c_3}q_k,\ \ \text{or}\ \ C\geq \left(\frac{8c_1}{c_3}\right)^2.$$
Squaring and summing \eqref{eq:Smlbb} over $m$ from $N_k/2$ to $N_k$, we obtain
\begin{align}
\sum_{m=1}^{N_k} |S_m(\omega)|^2\geq \frac{3c_1^2 C}{256} N_k^2>2N_k^2,
\end{align}
provided that 
$$C\geq \frac{512}{3 c_1^2}.$$
Finally it suffices to take 
\begin{equation}\label{def:constantC}
C=\max\left(16,\  \left(\frac{8c_1}{c_3}\right)^2,\ \frac{512}{3 c_1^2}\right).
\end{equation}
This proves Theorem \ref{thm:main3}.
\qed

\subsection{Proof of Lemma \ref{lm:Khinchin}}
An analogous lemma with the condition $2|q$ instead of $2|p$ can be found in \cite{FJK}. 
Here we need $2|p$.
Let us note that it suffices to show 
\begin{lm}\label{lm:Khinchin'}
Under the same condition as Lemma \ref{lm:Khinchin}, we have that for Lebesgue almost-every $\omega\in \mathbb{R}$, the inequality 
$$|q\omega-p|<\psi(q)$$
has infinitely many solutions $p,q$ with odd $q$ and $(p,q)=1$.
\end{lm}
First let us show how Lemma \ref{lm:Khinchin'} implies Lemma \ref{lm:Khinchin}. 
Indeed, Lemma \ref{lm:Khinchin'} with $\tilde{\psi}(n)=\psi(n)/2$ implies that, for Lebesgue almost-every $\omega$, there are infinitely many solutions $(p,q)$ to the inequality
\begin{align}\label{eq:Khinchin'}
\left|q\frac{\omega}{2}-p\right|<\frac{\psi(q)}{2},
\end{align}
with odd $q$ and $(p,q)=1$.
Multiplying \eqref{eq:Khinchin'} by $2$, we obtain
$$|q\omega-2p|<\psi(q),$$
where $(q, 2p)=(q,p)=1$.
This proves Lemma \ref{lm:Khinchin}.\\

Lemma \ref{lm:Khinchin'} is a special case of the following theorem in \cite{Harman}.
\begin{thm}\cite[Theorem 4.2]{Harman}\label{thm:Harman}
Suppose that $\psi(n)$ is a non-increasing sequence with $0<\psi(n)\leq 1/2$ and suppose that 
$$\sum_{n=1}^{\infty}\psi(n)=\infty.$$
Let $\mathcal{A}$ be an infinite set of positive integers. We write $\mathcal{S}(\mathcal{A}, \omega, N)$ for the number of solutions to 
$$\|n\omega\|_{\mathbb{T}}<\psi(n),\ \ n\leq N,\ \ n\in \mathcal{A},$$
where $\|x\|_{\mathbb{T}}:=\dist(x, \Z)$.
Then, for Lebesgue almost-every $\omega\in \mathbb{R}$, 
$$\mathcal{S}(\mathcal{A}, \omega, N)=2\Psi(N, \mathcal{A})+O(\Psi(N)^{1/2} (\log \Psi(N))^{2+\varepsilon}),$$
for every $\varepsilon>0$. Here
$$\Psi(N)=\sum_{n=1}^N \psi(n),\ \ \text{and}\ \ \Psi(N, \mathcal{A})=\sum_{\substack{n=1\\ n\in\mathcal{A}}}^N \psi(n).$$
\end{thm}

We can now give the

\be{proof}[Proof of Lemma \ref{lm:Khinchin'}]
Let us take $\mathcal{A}=\{2k-1,\ k\in \mathbb{N}\}$ be the set of positive odd numbers
Since $\psi(n)$ is a non-increasing sequence, we easily see that 
$$2\Psi(2N,\mathcal{A})\geq  \Psi(2N)\to \infty.$$
Thus Theorem \ref{thm:Harman} implies 
$$\mathcal{S}(\mathcal{A}, \omega, 2N)\to \infty.$$
This proves Lemma \ref{lm:Khinchin'}.
\e{proof}

\begin{appendix}
\section{The perturbation series in the Almost-Mathieu case}\label{app:AMO}
In this section only, we set $\tilde v_j=2\cos(2\pi(\om j+\theta))$. We use analogous notation as in the skew-shift case, occasionally using $\,\tilde{}\,$ for emphasis. By adapting the calculation in the proof of Proposition \ref{prop:pertseries}, we find that
the polynomial
$$
\tilde{\curly{P}}_n(\lam,0)
:=\mathrm{Tr}[\tilde M_n^* \tilde M_n]=\sum_{k=0}^{n} \tilde \al_{2k} \lam^{2k},
$$
now has coefficients given by $\tilde\al_0=2$ and
\beq\label{eq:tildeal2kdefn}
\tilde \al_{2k}=
\sum_{\substack{0\leq k_1,k_2\leq  n\\ k_1+k_2=2k} } (-1)^{\frac{k_1-k_2}{2}}
\sum_{\substack{1\leq j_1<\ldots<j_{k_1}\leq n\\ 1\leq l_1<\ldots<l_{k_2}\leq n}} \ind_{\clubsuit}(\ul{j},\ul{l})  \sum_{\substack{\ul{a}\in\{\pm 1\}^k\\ \ul{a}\perp\ul{1}}}  e\l[\om \sum_{s=1}^{k} a_s j_s\r]
\eeq
for $k\geq 1$.

Here we only consider the lowest non-trivial coefficient $\tilde \al_2$.

\be{prop}\label{prop:tilde}
We have 
\beq\label{eq:tildeal2}
\tilde\al_2=\frac{|1-e[(\om+\frac{1}{2}) n]|^2}{2\cos^2(\pi\om )}\leq \frac{2}{\cos^2(\pi\om )}.
\eeq
\e{prop}

We emphasize that the upper bound is independent of $n$. This is in stark contrast with the skew-shift model, for which $\al_2=2n$ was proved in Theorem \ref{thm:main}. This serves as an important indication that we have not lost the critical features of the models when applying Jensen's inequality to move from Conjecture \ref{conj} to \ref{conj'}.

\be{proof}[Proof of Proposition \ref{prop:tilde}]
This is a calculation. By definition, we have
$$
\begin{aligned}
\tilde\al_2
=&-4 \Re \sum_{1\leq j_1<j_{2}\leq n} \l(\ind_{(\ev,\od)}+\ind_{(\ev,\od)} \r)(j_1,j_2) e[\om(j_1-j_2)]\\
&+2\Re \sum_{1\leq j_1,l_1\leq n} \l(\ind_{(\ev,\ev)}+\ind_{(\od,\od)} \r)(j_1,l_1)  e[\om(j_1-j_2)]\\
=&2\Re \sum_{1\leq j_1,l_1\leq n} (-1)^{j_1+l_1}e[\om(j_1-j_2)]\\
=&2 \Re \l|\sum_{j=1}^n (-1)^j e[\om j]\r|^2\\
=&\frac{|1-e[(\om+\frac{1}{2}) n]|^2}{2\cos^2(\pi\om )},
 \end{aligned}
$$
as claimed.
\e{proof}

\section{The analysis of Weyl sums via homogeneous dynamics}
In recent years, Weyl-sum asymptotics have been established using techniques from homogeneous dynamics \cite{Cel,CM,Marklof}. In fact, it is possible to study the entire path traced out by the Weyl sum in the complex plane as a stochastic process in Wiener space (the space of continuous functions with $\sup$-norm). Notice that control on the entire path directly addresses the technical problem (ii) mentioned in Section \ref{ssect:weyl}.

In particular, Cellarosi-Marklof recently established that Weyl sums satisfy an ``invariance principle'': If the frequency $\om\in [0,1]$ is chosen at random, and $c_1\in \R$ is irrational, then the path of the normalized Weyl sum $m^{-1/2}\sum_{j=1}^{m} e\l[\om (j^2+c_1 j)\r]\in \C$ converges to a non-trivial random variable in Wiener space; see Theorem 1.3 in \cite{CM}. The limit shares some properties with two-dimensional Brownian motion, but is distinct from it; see Theorem 1.4 in \cite{CM}. The fact that the parameter $c_1$ is \emph{irrational} guarantees the validity of the key equidistribution theorems for horocycles in an appropriate hyperbolic space. Note that, for us, $c_1=-1$ is rational and so their theorems do not apply in our context. (In fact, we might expect that, instead of equidistribution in the whole space, one now has equidistribution along a certain geodesic.)

Nonetheless, in this appendix, we observe here that the tightness part of their proof extends to our situation. We recall that $S_{m}= \sum_{j=1}^{m} e\l[\om (j^2-j)\r]$ and define the function $X_{N}:[0,1]\to \C$ by
\beq\label{eq:XNdefn}
\begin{aligned}
X_{n}(t):=& n^{-1/2}\l(S_{\floor{tn}}+(tn-\floor{tn}) (S_{\ceil{tn}}-S_{\floor{tn}})\r).
\end{aligned}
\eeq
In other words, we take $X_{n}(t):= n^{-1/2}S_{tn}$ whenever $t\in \l\{0,\frac{1}{n},\frac{2}{n},\ldots,1\r\}$, and we interpolate linearly between these points. We now sample the frequency $\om\in [0,1]$ at random, according to a fixed measure $\mu$ that is absolutely continuous with respect to Lebesgue measure. The random choice of $\om$ induces a stochastic process $X_{N}$ in the Wiener space $\curly{C}$, defined as the Banach space $C([0,1];\C)$ equipped with the supremum norm. 

The following tightness result is implicit in \cite{CM}.

\be{thm}\label{thm:compactness}
The sequence of stochastic processes $(X_n)_{n\geq 1}$ is pre-compact under weak convergence in Wiener space $\curly{C}$.
\e{thm}

\be{proof}
To see this, it suffices to observe that the proof of Proposition 4.10 of \cite{CM} extends verbatim to sums over $e[\om(j^2-j)]$. In particular, the tail bounds in Proposition 3.17 of \cite{CM} hold uniformly in the vector $\mathbf{\xi}$ and do not require the irrationality assumption stated in Theorem 1.3 there.
\e{proof}

We can use Theorem \ref{thm:compactness} to obtain a limiting continuous random process $X_0:[0,1]\to \C$ such that we have weak convergence in Wiener space, $X_{n}\stackrel{d}{\to} X_0$, along a subsequence. We emphasize that the limit $X_0$ can be different from the limiting process $X$ found in \cite{CM} under an additional irrationality assumption. Nonetheless, the two limiting processes are likely related. See also Remark 1.1 in \cite{CM}.

\end{appendix}


\begin{thebibliography}{99}

\bibitem[AizWar]{AW}
Aizenman, M., Warzel, S. {\em Random operators. 
Disorder effects on quantum spectra and dynamics}, Graduate Studies in Mathematics, 168. American Mathematical Society, Providence, RI, 2015

\bibitem[And]{Anderson}
Anderson, P.W. {\em Absence of diffusion in certain random lattices}, Phys.\ Rev.\ 109 (1958)  (5): 1492--1505. 

\bibitem[BelSim]{BelSim}  B\'ellissard, J., Simon, B.  {\em 
Cantor spectrum for the almost Mathieu equation},
J.\ Funct.\ Anal.\ 48 (1982), no.~3, 408--419.

\bibitem[BerGol]{BerryGoldberg}
Berry, M.V., Goldberg, J. {\em Renormalisation of curlicues}, Nonlinearity 1 (1988), no.~1, 1--26. 

\bibitem[Ble]{Ble}
Bleher, P.M., {\em The energy level spacing for two harmonic oscillators with generic ratio of frequencies}, J.\ Stat.\ Phys.\ 63 (1991), no.~1 and 2, 261--283. 

\bibitem[Bou1]{B1}   
Bourgain, J. {\em Green's function estimates for lattice Schr\"odinger operators and applications}, Annals of Mathematics Studies, 158. 
Princeton University Press, Princeton, NJ, 2005. 
 
\bibitem[Bou2]{B2}  
Bourgain, J. {\em 
On the spectrum of lattice Schr\"odinger operators with deterministic potential},
Dedicated to the memory of Thomas H. Wolff. 
J.\ Anal.\ Math.\ 87 (2002), 37--75. 

\bibitem[BouGol]{BG} Bourgain, J., Goldstein, M. {\em On nonperturbative localization with quasi-periodic potential},  Ann.\ of Math.\ (2) 152 (2000), no.~3, 835--879.

\bibitem[BouGolSch]{BGS} Bourgain, J.,  Goldstein, M.,  Schlag, W.  {\em Anderson localization for Schr\"odinger operators on $\Z$ with potentials 
 given by the skew-shift},  Comm.\ Math.\ Phys.\ 220 (2001), no.~3, 583--621. 

\bibitem[BouJit]{BJ}
Bourgain, J., Jitomirskaya, S., {\em Continuity of the Lyapunov Exponent for Quasiperiodic Operators with Analytic Potential}, J.\ Stat.\ Phys.\ 108 (2002), no. 5-6,  1203–-1218

\bibitem[Cel]{Cel}
 Cellarosi, F. {\em Limiting curlicue measures for theta sums},  Ann.\ Inst.\ Henri Poincar\'e Probab.\ Stat.\ 47 (2011), no.~2, 466--497. 

\bibitem[CelMar]{CM}
Cellarosi, F., Marklof, J.
{\em Quadratic Weyl sums, automorphic functions and invariance principles},
Proc.\ Lond.\ Math.\ Soc.\ (3) 113 (2016), no.~6, 775--828. 

\bibitem[ChuSp]{CS}
Chulaevsky, V., Spencer, T.,
{\em Positive Lyapunov exponents for a class of deterministic potentials},
Comm.\ Math.\ Phys.\ 168 (1995), no.~3, 455--466. 

\bibitem[CouKaz]{CK}
Coutsias, E.A., Kazarinoff, N.D. {\em The approximate functional formula for the theta function and Diophantine Gauss sums}, Trans.\ Amer.\ Math.\ Soc.\ 350 (1998), no. 2, 615--641. 

\bibitem[FedKlo]{FK}
Fedotov, A., Klopp, F., {\em An exact renormalization formula for Gaussian exponential sums and applications},    Amer.\ J.\ Math.\ 134 (2012), no.~3, 711--748. 

\bibitem[FieJuK{\"o}r]{FJK}
Fiedler, H., Jurkat, W., K{\"o}rner, O., {\em Asymptotic expansions of finite theta series}    Acta Arith.\ 32 (1977),   no.~2, 129--146. 

\bibitem[FigPas]{FigPas} Figotin, A., Pastur, L., {\em Spectra of random and almost-periodic operators}, Grundlehren der Mathematischen Wissenschaften, 297. Springer-Verlag, Berlin, 1992.

\bibitem[Fur]{Fur}   F\"{u}rstenberg, H. {\em Noncommuting random products}, Trans.\ Amer.\ Math.\ Soc.~108 (1963), 377--428.

\bibitem[GolSch]{GS}  Goldstein, M.,  Schlag, W. {\em H\"older continuity of the integrated density of states for quasi-periodic Schr\"odinger equations and averages of shifts of subharmonic functions}, Ann.\ of Math.\ (2) 154 (2001), no.~1, 155--203.

\bibitem[HanLemSch]{HLS} Han, R., Lemm, M., Schlag, W. {\em Effective multi-scale approach to the Schr\"odinger cocycle over a skew shift base}, 	arXiv:1803.02034

\bibitem[Har]{Harman} Harman, G., {\em Metric number theory}, Courier Corporation (1998).


\bibitem[H-B]{DRH} Heath-Brown, D.\ R. {\em 
Pair correlation for fractional parts of $\alpha n^{2}$}, 
Math.\ Proc.\ Cambridge Philos.\ Soc.~148 (2010), no.~3, 385--407.

\bibitem[HarLit]{HL}
Hardy, G.H., Littlewood, J.E. {\em Some problems of diophantine approximation}, Acta Math.\ 37 (1914), no.~1, 193--239. 

\bibitem[Her]{Her}   Herman, M.-R. {\em Une m\'ethode pour minorer les exposants de Lyapounov et quelques exemples montrant le caract\`ere local d'un th\'eor\`eme d'Arnol'd et de Moser sur le tore de dimension 2},  Comment.\ Math.\ Helv.\ 58 (1983), no.~3, 453--502. 

\bibitem[JuVHo]{JvH}
Jurkat, W. B., Van Horne, J. W.,
{\em The proof of the central limit theorem for theta sums}, 
Duke Math.\ J.\ 48 (1981), no.~4, 873--885.

\bibitem[Kru]{Kru}
Kr{\"u}ger, H., {\em An explicit skew-shift Schr\" odinger operator with positive Lyapunov exponent at small coupling}, 	arXiv:1206.1362

\bibitem[KunSou]{KS}
Kunz, H., Souillard, B. {\em Sur le spectre des op{\'e}rateurs aux diff{\'e}rences finies al{\'e}atoires}, Comm.\ Math.\ Phys.\ 78 (1980/81), no.~2, 201--246.

\bibitem[Leh]{Lehmer}
Lehmer, D.\ H.,
{\em Incomplete Gauss sums},
Mathematika 23 (1976), no.~2, 125--135. 

\bibitem[Mar]{Marklof}
Marklof, J., {\em Limit theorems for theta sums}, Duke Math.\ J.\ 97 (1999), no.~1, 127--153. 

\bibitem[MarStr]{MS} Marklof, J., Str\"ombergsson, A. {\em 
Equidistribution of Kronecker sequences along closed horocycles},  
Geom.\ Funct.\ Anal.~13 (2003), no.~6, 1239--1280. 

\bibitem[MarYe]{MY}
Marklof, J., Yesha, N. {\em Pair correlation for quadratic polynomials mod 1}, Compos.\ Math.\ 154 (2018), no.~5, 960--983. 

\bibitem[Mon]{Mont} Montgomery, Hugh L. {\em Ten lectures on the interface between analytic number theory and harmonic analysis.} CBMS Regional Conference Series in Mathematics, 84.  American Mathematical Society, Providence, RI, 1994.

\bibitem[Mor]{Mordell}
Mordell, L.\ J., 
{\em The Approximate Functional Formula for the Theta Function},
J.\ London Math.\ Soc.\ 1 (1926), no.~2, 68--72. 

\bibitem[PanBoGia]{PBG}
Pandey, A., Bohigas, O., Giannoni, M.J., {\em Level repulsion in the spectrum of two-dimensional harmonic oscillators}, 
J.\ Phys.\ A: Math.\ Gen.\ 22 (1989), no.~18, 4083--4088.

\bibitem[RudSar]{RS}
Rudnick, Z.,  Sarnak, P.,
{\em The pair correlation function of fractional parts of polynomials}, 
Comm.\ Math.\ Phys.\ 194 (1998), no.~1, 61--70. 

\bibitem[RudSarZah]{RSZ}   Rudnick, Z.,  Sarnak, P.,  Zaharescu, A. {\em The distribution of spacings between the fractional parts of $n^{2}\alpha$},  Invent.\ Math.\ 145 (2001), no.~1, 37--57.

\bibitem[SorSpe]{SoSp} Sorets, E.,  Spencer, T. {\em Positive Lyapunov exponents for Schr\"{o}dinger operators with quasi-periodic potentials},
Comm.\ Math.\ Phys.\ 142 (1991), no.~3, 543--566. 

\bibitem[Via]{Viana}  Viana, M. {\em Lectures on Lyapunov exponents}, Cambridge Studies in Advanced Mathematics, 145. Cambridge University Press, Cambridge, 2014.

\bibitem[Wey1]{W1}
Weyl, H. {\em {\"U}ber ein Problem aus dem Gebiete der diophantischen Approximationen}, Nachr.\ Ges.\ Wiss.\ G\"{o}ttingen, Math.-Phys.\ Kl.\, 234--244, 1914. Reprinted in Gesammelte Abhandlungen, Band I. Berlin: Springer-Verlag, pp. 487--497, 1968

\bibitem[Wey2]{W2}
Weyl, H. {\em {\"U}ber die Gleichverteilung von Zahlen mod.\ Eins}, Math.\ Ann.\ 77, 313-352, 1916. Reprinted in Gesammelte Abhandlungen, Band I. Berlin: Springer-Verlag, pp. 563--599, 1968

\bibitem[Wil]{Wilton}
Wilton, J.R., {\em The Approximate Functional Formula for the Theta Function}, J.\ London Math.\ Soc.\ 2 (1927), no.~3, 177--180

\end{thebibliography}
\end{document}